\documentclass[aps,pra,longbibliography,reprint,showpacs,superscriptaddress,groupedaddress]{revtex4-2}  
\usepackage{graphicx}	
\usepackage{dcolumn}	
\usepackage{bm}		
\usepackage{amssymb}	
\usepackage{amsmath}	
\usepackage{multirow}	
\usepackage{array}

\newcolumntype{C}{@{\extracolsep{0.6cm}}c@{\extracolsep{0.4cm}}}

\begin{document}
\title{Correlation-Driven Phenomena in Periodic Molecular Systems from Variational Two-electron Reduced Density Matrix Theory}

\author{Simon Ewing}
\affiliation{Department of Chemistry and The James Frank Institute, The University of Chicago, Chicago USA}
\author{David A. Mazziotti}
\email{damazz@uchicago.edu}
\affiliation{Department of Chemistry and The James Frank Institute, The University of Chicago, Chicago USA}

\date{Submitted March 13, 2021}

\begin{abstract}
Correlation-driven phenomena in molecular periodic systems are challenging to predict computationally not only because such systems are periodically infinite but also because they are typically strongly correlated.   Here we generalize the variational two-electron reduced density matrix (2-RDM) theory to compute the energies and properties of strongly correlated periodic systems.   The 2-RDM of the unit cell is directly computed subject to necessary $N$-representability conditions such that the unit-cell 2-RDM represents at least one $N$-electron density matrix.   Two canonical but non-trivial systems,  periodic metallic hydrogen chains and periodic acenes, are treated to demonstrate the methodology.   We show that, while single-reference correlation theories do not capture the strong (static) correlation effects in either of these molecular systems, the periodic variational 2-RDM theory predicts the Mott metal-to-insulator transition in the hydrogen chains and the length-dependent polyradical formation in acenes.   For both hydrogen chains and acenes the periodic calculations are compared with previous non-periodic calculations with the results showing a significant change in energies and increase in the electron correlation from the periodic boundary conditions.  The 2-RDM theory, which allows for much larger active spaces than are traditionally possible, is applicable to studying correlation-driven phenomena in general periodic molecular solids and materials.
\end{abstract}

\maketitle
\section{\label{sec:Intro}Introduction}

Computing the electronic structure of extended molecules and materials can reveal important information such as the band gap, reactivity, and bulk properties such as conductivity or polarizability. As such, accurate methods for computing the electronic structure of materials is of interest to many fields, including the study of inorganic polymers, organic electronics, semiconductors, and superconductors~\cite{Andre1984, Hachmann2007, Dimitrakopoulos2002, Reese2004, Chen2018, Kaewmeechai2019, Kim2011, Zerveas2016, Xie2020, Benayad2014, Liu2015, Safaei2018}.  Electronic structure calculations on extended materials are only computationally tractable in cases where periodic boundary conditions can be imposed, which separate the electrons and orbitals into smaller, periodically repeating, unit cells. Commonly used methods for such calculations include density functional theory (DFT)~\cite{Hohenberg1964, Kohn1965, Enkovaara2010, Sundararaman2013}, GW approximations~\cite{Hedin1965}, quantum Monte Carlo methods~\cite{Caffarel1988}, and coupled cluster (CC) theory~\cite{Coester1958}.  Many of these methods, however, have difficulty computing strongly correlated periodic materials with high accuracy at an efficient computational cost, and hence, there is a need for further advances in methods and theories.

Here we present an approach to the calculation of electronic structures for periodic materials in the gamma-point approximation based on the variational calculation of the 2-electron reduced density matrix (2-RDM).  In the variational 2-RDM (v2RDM) method~\cite{McWeeny1960, Nakata2001, Mazziotti2002, Zhao2004, Gidofalvi2005, Cances2006, MazziottiBook2007, Gidofalvi2008, Greenman2010, Shenvi2010, Verstichel2012, MazziottiChemRev2012, MazziottiPRL2016, Fosso-Tande2016}, the 2-RDM is constrained by $N$-representability conditions that are necessary for the 2-RDM to represent at least one $N$-electron density matrix.  Because these conditions are necessary, the minimization of the energy with respect to the 2-RDM generates a lower bound on the ground-state energy in the given basis set.  Furthermore, because the constraints known as $p$-positivity conditions restrict the metric matrices of  $q$ particles and $(p-q)$ holes to be positive semidefinite~\cite{Mazziotti2001, MazziottiPRL2012}, the variational 2-RDM method has a well-defined, physical solution even in the presence of strong electron correlation.  The 2-positivity conditions, which include the nonnegativity of the particle-particle $^{2} D$, hole-hole $^{2} Q$, and particle-hole $^{2} G$ matrices, have been shown to be capable of describing strongly correlated phenomena including polyradical character in extended conjugated systems~\cite{Gidofalvi2008, Hemmatiyan2019}, ligand non-innocence in transition-metal complexes~\cite{Schlimgen2016, McIsaac2017}, non-superexchange mechanisms in bridged transition-metal dimers~\cite{Boyn2020},  and exciton condensation in electron double layers~\cite{Safaei2018}.

Extension of the v2RDM theory to treat periodic boundary conditions allows us to treat the electronic structure of strongly correlated periodic molecules without performing multiple molecular (open boundary condition) calculations and extrapolating to the infinite-length limit.  Because the periodic v2RDM method with the 2-positivity conditions only scales as $\mathcal{O}(r^6)$ where $r$ is the number of active orbitals in the unit cell, we can perform calculations with much larger active spaces than possible with traditional methods~\cite{Schlimgen2016, Montgomery2018}.  Through the $N$-representability conditions the v2RDM theory includes multi-body correlation that are difficult for many traditional periodic methods to capture.  Although we present the periodic v2RDM method in the gamma-point approximation here, the formalism can be extended to include $k$-point sampling.  To demonstrate the periodic v2RDM theory, we treat two extended systems, known for exhibiting strong electron correlation: hydrogen chains and acene chains.  Both systems are difficult to treat accurately with traditional methods like second-order many-body perturbation theory, configuration interaction with single and double excitations, or coupled cluster theory with single and double excitations extended to periodic boundary conditions.  The hydrogen chains undergo a Mott metal-to-insulator transition~\cite{Bendazzoli2011, Motta2019} upon dissociation with the insulator phase being strongly correlated, whereas the acene chains become strongly correlated with polyradical character as the lengths of the chains increase.

\section{\label{sec:App}Theory}

We discuss the v2RDM theory, periodic boundary conditions, and their combination into a periodic v2RDM method.

\subsection{\label{methods:v2RDM}Variational 2-electron Reduced Density Matrix Methods}

The 1-RDM and 2-RDM are defined by integrating the full $N$-electron density matrix over the spatial and spin coordinates of all but one or two electrons, respectively. Using second quantization, we can represent the elements of the 1- and 2-RDMs as
\begin{equation}\label{Eq:1RDM}
{}^1D^i_j=\langle\Psi|\hat{a}^\dagger_i\hat{a}_j|\Psi\rangle
\end{equation}
\begin{equation}\label{Eq:2RDM}
{}^2D^{i,j}_{k,l}=\frac{1}{2}\langle\Psi|\hat{a}^\dagger_i\hat{a}^\dagger_j\hat{a}_l\hat{a}_k|\Psi\rangle
\end{equation}
where $\hat{a}^\dagger_m$ and $\hat{a}_m$ are the second-quantized creation and annihilation operators for spin orbital $|\psi_m\rangle$~\cite{Coleman1963, Garrod1964, Mazziotti2002, MazziottiBook2007, Nakata2001, McWeeny1960, Zhao2004}. Notably, the 1-RDM can be derived from the 2-RDM by integrating over the spatial and spin coordinates for one of the two electrons. These spin RDMs can be converted into spatial RDMs by tracing over the spin of the electrons. Eigenvalues of spatial 1-RDMs, also known as spatial natural orbital occupation numbers~\cite{Lowdin1956}, represent the numbers of electrons in the spatial orbitals, and by the Pauli exclusion principle they are constrained to lie between 0 and 2.  A signature of strong correlation, or contributions from multiple Slater determinants, is the presence of fractionally filled orbitals, which are characterized by eigenvalues of the spatial 1-RDM near one~\cite{Head-Gordon2003}. Similarly, eigenvalues of 2-RDMs represent the number of electrons in each two-electron function, known as a geminal, and partially-filled geminals contribute to correlation.

For systems that have at most pairwise interactions, the molecular energy can be written as a functional of the 2-electron reduced density matrix:
\begin{equation}\label{Eq:energy}
E=\text{Tr}\left({}^1H{}^1D\right) + \text{Tr}\left({}^2V{}^2D\right)
\end{equation}
where ${}^1H$ and ${}^2V$ are the reduced Hamiltonian matrices of the 1-electron and 2-electron integrals. The variational 2-RDM (v2RDM) method minimizes the ground-state energy in Eq.~\ref{Eq:energy}, with the 2-RDM as the fundamental variable rather than the wavefunction~\cite{Gidofalvi2005, Mazziotti2002, MazziottiPRL2011}. Often, the v2RDM method is combined with the notion of an active space, a set of orbitals that are correlated in a mean field of the remaining orbitals~\cite{Gidofalvi2008, Roos1980}. A system where $n$ electrons are allowed to fill $r$ spatial orbitals is denoted an $[n,r]$ active space. If $n$ is equal to the total number of electrons and $r$ is equal to the size of the basis set, then the calculation approximates the energies from full configuration interaction (FCI).

Central to RDM calculations is the concept of $N$-representability, which requires that an RDM represents a physical $N$-electron density matrix~\cite{Coleman1963, Mazziotti2006, MazziottiPRL2012}. The simplest and most familiar $N$-representability constraint is the Pauli exclusion principle, which states that eigenvalues of the spatial 1-RDM must lie between 0 and 2. Similar constraints exist for the 2-electron RDM, the particle-hole RDM (${}^2G$), and the 2-hole RDM (${}^2Q$), namely ${}^2D\succeq0$, ${}^2G\succeq0$, and ${}^2Q\succeq0$, restricting the eigenvalues of all three matrices to be non-negative. All three of these constraints restrict the space of valid 2-RDMs because ${}^2G$ and ${}^2Q$ are related by linear mappings to the 2-RDM~\cite{Mazziotti2002}. Minimizing the energy from a 2-RDM subject to these constraints generates an optimization problem known as a semidefinite program~\cite{Vandenberghe1996, MazziottiPRL2004, MazziottiESAIM2007, MazziottiPRL2011}. The solution of the semidefinite program yields a lower bound to the ground-state energy from a complete active space configuration interaction (CASCI) calculation~\cite{Roos1980} but with a computational scaling that is polynomial in the size of the active space.

\subsection{\label{methods:PBCs}Periodic Boundary Conditions}

For molecules with extended structures, additional symmetries must be exploited to make accurate electronic structure calculations tractable. Calculations on periodic molecules can be altered to account for the periodic boundary conditions (PBCs) of the molecule by using periodic basis functions. Bloch waves are the simplest way to construct a  complete basis set of periodic functions, but the underlying structure for these basis functions can vary, and are usually chosen to be either Gaussian~\cite{Soler2002, Sun2018} or plane waves~\cite{Giannozzi2009, Gonze2002, Kresse1996, Payne1992}. In general, Bloch waves can be defined as
\begin{equation}\label{Eq:bloch}
|\Psi_k(r)\rangle=e^{ik\cdot r}u(r)
\end{equation}
where $u(r)$ is a function for which the periodic boundary conditions are satisfied and $k$ is a momentum vector. For many cases, the gamma point, where only $k=0$ wave vectors are included in the basis set, gives an accurate approximation to the complete basis while greatly simplifying the computational complexity. Although plane waves automatically satisfy the periodic boundary conditions, they do not resolve atomic details as readily as Gaussian basis functions. In the case of Gaussian functions periodic boundary conditions can be imposed through a local basis approximation~\cite{Sun2018}:
\begin{equation}\label{Eq:periodic}
|\Psi_k(r)\rangle=\sum_Te^{ik\cdot T}u(r-T)
\end{equation}
where $T$ is a lattice translational vector, and $u(r-T)$ is a local Gaussian atomic basis function. This approximation accounts for periodicity by summing over images of each basis function in neighboring cells. As the number of translational vectors in the sum over $T$ increases, the wavefunction will more closely approach the periodic boundary conditions, because the closest cells are being explicitly included. Once the basis set is chosen to satisfy the boundary conditions, the 1-electron and 2-electron integral matrices can be computed and used in the v2RDM method. We use the PySCF built-in integration methods to evaluate these integrals using the local basis approximation.

Without the use of periodic boundary conditions, extended systems are typically approximated by computing molecular systems of varying sizes and extrapolating to the infinite limit. In this form of analysis, several assumptions are made about the energy density of the system so that the energies of the different systems can be subtracted, leaving an ``effectively edgeless" system. Namely, it is assumed that the system is long enough that the system can be broken into several classes of edge subsystems and edgeless (quasi-periodic) subsystems. By subtracting the energy of smaller systems that contain no (or fewer) edgeless subsystems, the energy of the central systems can be approximated. However, gamma point calculations compute these systems directly due to the PBCs used in the unit cell, and these calculations are often much cheaper because there is no need to compute the electronic structure of a supercell. Additionally, we posit that gamma point calculations better represent the extended systems than molecular calculations of the same size, for two main reasons: first, molecular calculations typically require several calculations to get an energy so that the analysis above can be performed. Second, periodic calculations eliminate the computational time and power needed to compute the electronic structure of the edge subsystems, which are ultimately discarded. Additionally, this discarding of data unnecessarily complicates the analysis of these computations for obervables other than the energy.

In both forms of analysis, the assumption that the (quasi) periodic unit cells are identical - no phase change between cells - can be relaxed to get a more accurate representation of the extended structure, since long-range order can affect the electronic structure in extended systems. For molecular calculations, the analysis above is simply extended further to larger systems, which reduces the inaccuracy due to each approximation. Using PBCs, there are two main methods for relaxing this assumption: including $k$-points or including more repeating units in the unit cell. Including $k$-points explicitly adds basis functions that allow for phase changes between unit cells. Similarly, including more repeating units in the unit cell allows the wavefunction to change phase between repeating units within the cell. Although $k$-point calculations allow sampling of the Brillouin zone, adding repeating units to the unit cell reduces the size of the Brillouin zone, meaning that less sampling is needed. We chose to use this method to analyze our systems due to the computational ease, and we will examine the effects of $k$-point calculations on strong correlation in the future.

\begin{figure*}[t]
\includegraphics[scale=0.25]{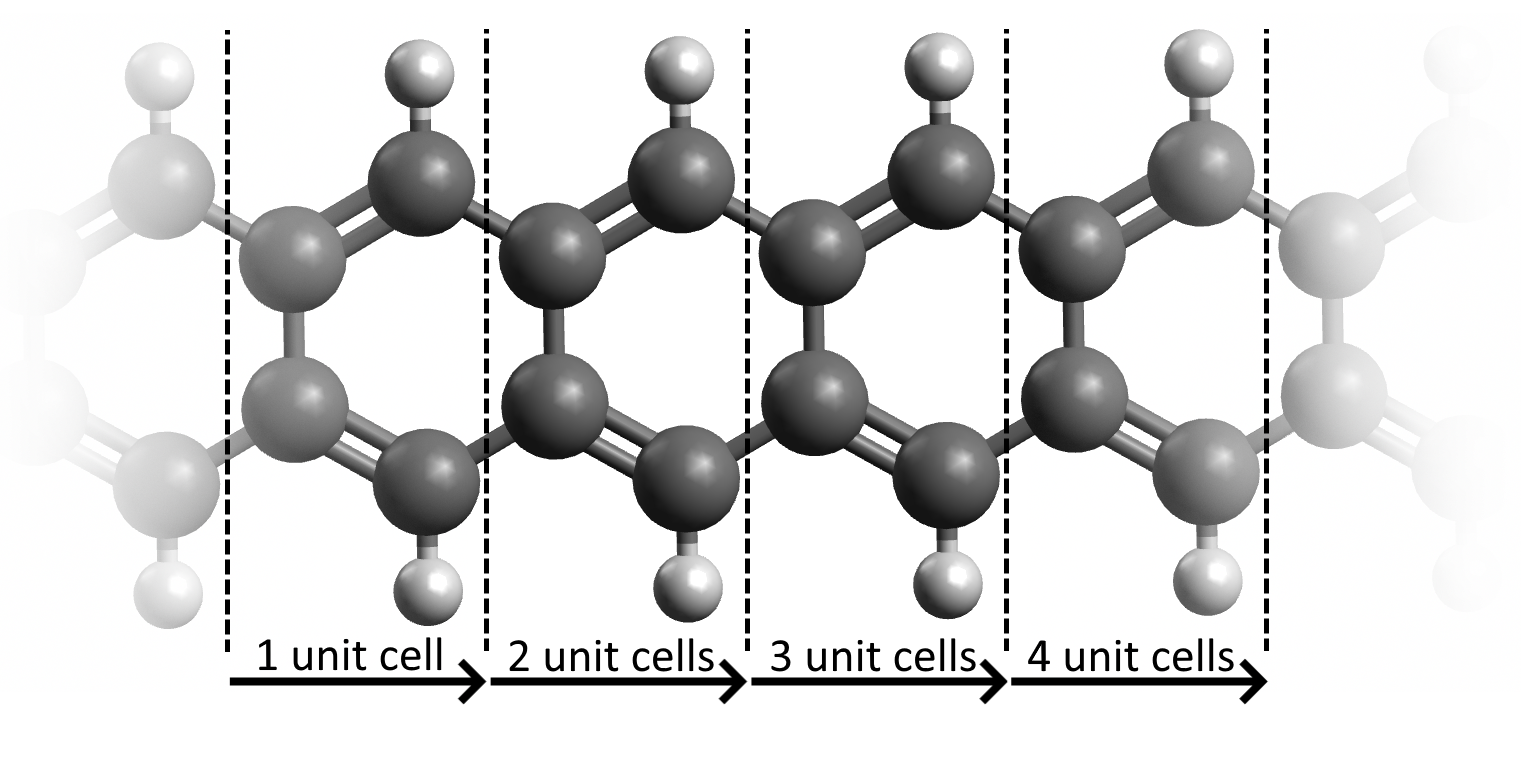}
\caption{The acene chain systems used for calculations have various numbers of repetitions of the unit cell inside one periodic box. Every unit cell contains 4 carbon atoms and 2 H atoms, for a total of 26 electrons and 40 orbitals in the 6-31G basis set. Of those orbitals, 4 electrons and 4 orbitals ([4,4] active space) were used from each unit cell in the active space calculations.}\label{Fig:Acene_graphic}
\end{figure*}

\section{\label{sec:Results}Results}

\subsection{\label{App:Methods}Methods}

Hydrogen chain symmetric dissociation curves were computed with $\textup H_{10}$ in the unit cell, using the correlation-consistent polarized valence double-zeta (cc-pVDZ)~\cite{Dunning1989} basis set. Both molecular and periodic calculations were completed in a [10,20] active space. Additional calculations were performed using DFT (B3LYP functional), configuration interaction with single and double excitations (CISD), and M{\o}ller-Plesset 2nd-order perturbation theory (MP2)~\cite{Moller1934} methods.  Each of these calculations were performed using the implementations in PySCF.

Additionally, the electronic structure of acene chains were computed using the periodic v2RDM method and the 6-31g basis set, with crystal structure coordinates obtained from the American Mineralogist Crystal Structure Database~\cite{Downs2003} and explicitly shown in the Supporting Information Table S2. A series of calculations were performed, with 1-10 repetitions of the unit cell inside the periodic box. For each calculation, we used a [4,4] active space per repeated unit, which accounts for the complete $\pi$-space. Therefore, the 10-unit calculation had an active space of [40,40]. This procedure ensured that the energy per repeated unit and occupation numbers were converged, and revealed information about the parity of the periodic orbitals. The occupation numbers from periodic v2RDM were then compared to those obtained via gamma-point calculations using PySCF-builtin periodic coupled cluster methods~\cite{Sun2018}. Finally, the energy, natural orbital (NO) occupation numbers, and several forms of entropy were computed for molecular acene chains with 2, 4, 6, and 8 rings, using the v2RDM method. The molecular calculations had active spaces of [10,10], [18,18], [26,26], and [34,34], respectively, corresponding once again to the full $\pi$-space. Geometries for these molecular chains were obtained from the supplementary data from Ref.~\cite{Hachmann2007}. These results are compared to the calculations using the periodic method.

\subsection{\label{Res:hydrogen}Hydrogen Chain}

\begin{figure}[t]
\includegraphics[scale=0.5]{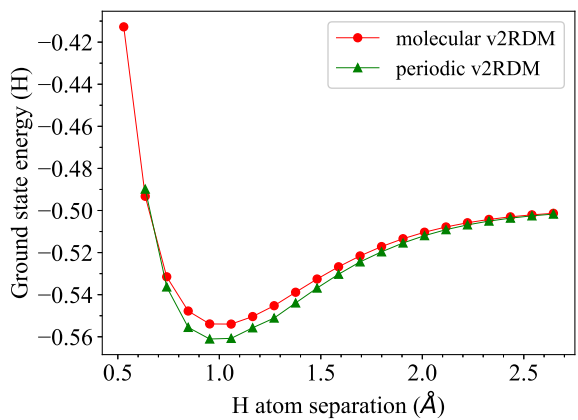}
\caption{Dissociation curves for $\textup H_{10}$, using molecular v2RDM and periodic v2RDM with a [10,20] active space and the cc-pVDZ basis set. Results agree with previous data~\cite{Stella2011, Motta2017}, showing the equilibrium bond length at 0.95-1.05 \AA.  A comparison to a periodic CASCI method is presented in the Supporting Information Table~S1.}\label{Fig:H_dissoc}
\end{figure}

Previous work on the extended hydrogen chain has focused on obtaining the dissociation curve through various extrapolation techniques, starting from molecular calculations~\cite{Motta2017, fripiat1983}. Molecular energies for multiple chain lengths are calculated, and then compared with each other to extrapolate to the infinite chain length, or the thermodynamic, limit.  We use the PBC approach to compute the ground state energy of the hydrogen chain, which removes the edge effects entirely without extrapolation. To verify the periodic-system-to-molecule limit, we present ground-state energies for 1-, 2-, and 3-dimensional hydrogen clusters with varying periodic box sizes in Supporting Information Figure~S1.

Examining the dissociation curve for $\textup H_{10}$ in Fig.~\ref{Fig:H_dissoc} we observe that the molecular and periodic systems converge to the same energy, around 5~hartrees, as the spacing between hydrogen atoms increases.  In this regime, the hydrogen atoms begin to behave independently, so the total energy is approximately 10 times the energy of a single hydrogen atom. However, interesting differences develop as the separation distance decreases. The equilibrium bond length lies at 0.95-1.05 \AA, and the periodic system has a deeper well by 6.6 millihartrees per atom. This agrees with previous work by Motta \textit{et al}.~\cite{Motta2017} that indicates through extrapolating molecular calculations to the infinite limit that the periodic system has a deeper well by about 4 millihartrees per atom.

\begin{figure}
\includegraphics[scale=0.5]{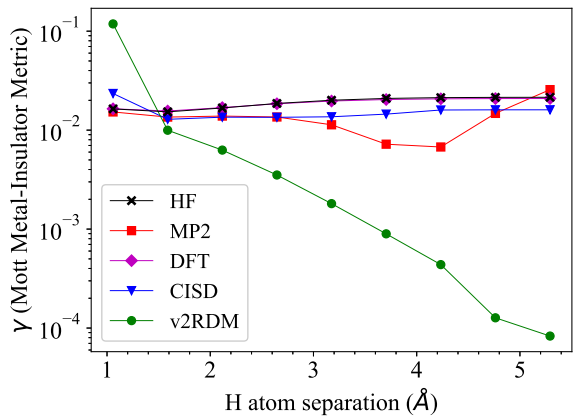}
\caption{Mott metal-to-insulator metric as a function of hydrogen atom spacing for $\textup H_{10}$, for various periodic methods. Hartree-Fock and other post-Hartree-Fock calculations produce a metric that is relatively constant across all bond lengths, whereas both molecular~\cite{Sinitskiy2010} and periodic v2RDM calculations have an appropriate transition from metal to insulator as the bond length increases.}\label{Fig:Mott}
\end{figure}

One metric that has been used~\cite{Sinitskiy2010,Smart2019,Tsuchimochi2009} to discuss the Mott metal-to-insulator transition is the sum of the magnitudes of the off-diagonal elements of the 1-RDM in the atomic orbital basis, denoted $\gamma$. Because the metal-to-insulator transition is a long-range effect, we ignore elements corresponding to orbitals on a single atom, instead summing over the off-diagonal elements that correspond to electron density shared between atoms. We add these elements of the 1-RDM in quadrature, as a parallel to the Frobenius norm of the matrix. See the Supporting Information for additional details. When the system has long-range order, this metric will be large, indicating strong metallic behaviour, and when the system has no long-range order, this metric will be small, indicating strong insulating behaviour.  Fig.~\ref{Fig:Mott} shows the metric as a function of the atomic spacing in $\textup H_{10}$.  The Mott transition metric for periodic Hartree-Fock calculations is nearly constant for all bond lengths. A similar absence of the transition to an insulator is seen from the DFT with the B3LYP functional, MP2, and CISD (As observed in previous work~\cite{MazziottiPRL2004}, the coupled cluster calculations with single and double excitations do not converge far beyond the equilibrium bond length, and hence, it is not included in the reported data).  By contrast, the periodic v2RDM calculations exhibit the expected transition from metallic to insulating properties as the bond length increases.

\subsection{\label{Res:acene}Acene Chain}

\begin{table*}
\begin{ruledtabular}
\begin{tabular}{l*{4}{Cc}}
	       & \multicolumn{8}{c}{Natural Orbital Occupation Numbers}\\
	\cline{2-9}
	       & \multicolumn{2}{c}{1 unit cell} & \multicolumn{2}{c}{2 unit cells} & \multicolumn{2}{c}{3 unit cells} & \multicolumn{2}{c}{4 unit cells}\\
	\cline{2-3}\cline{4-5}\cline{6-7}\cline{8-9}
	       & CCSD & v2RDM & CCSD & v2RDM & CCSD & v2RDM & CCSD & v2RDM\\
	\hline
	HONO-2 & 1.9594 &      -      & 1.9082 & 1.9036 & 1.9361 & 1.9321 & 1.8992 & 1.8662 \\
	HONO-1 & 1.9580 & 1.9842 & 1.9055 & 1.8918 & 1.9032 & 1.8515 & 1.8992 & 1.8662 \\
	HONO    & 1.9440 & 1.9743 & 1.4740 & 1.3558 & 1.9032 & 1.8515 & 1.7194 & 1.3479 \\
	LUNO     & 0.0475 & 0.0263 & 0.5215 & 0.6457 & 0.0902 & 0.1511 & 0.2782 & 0.6597 \\
	LUNO+1 & 0.0391 & 0.0152 & 0.0845 & 0.1109 & 0.0902 & 0.1511 & 0.0922 & 0.1309 \\
	LUNO+2 & 0.0301 &      -      & 0.0841 & 0.0943 & 0.0548 & 0.0688 & 0.0922 & 0.1309 \\
\end{tabular}
\end{ruledtabular}
\caption{Natural orbital occupations for acene chains with 1-4 unit cells in the periodic box.  v2RDM calculations show strong correlation for even numbers of unit cells, whereas CCSD fails to recover strong correlation except in the 2 unit cell calculation. CCSD calculations were only performed up to 4 unit cells due to the expensive memory requirements for longer chains.}\label{Tab:Acene_occ}
\end{table*}

\begin{figure}[t]
\includegraphics[scale=0.38]{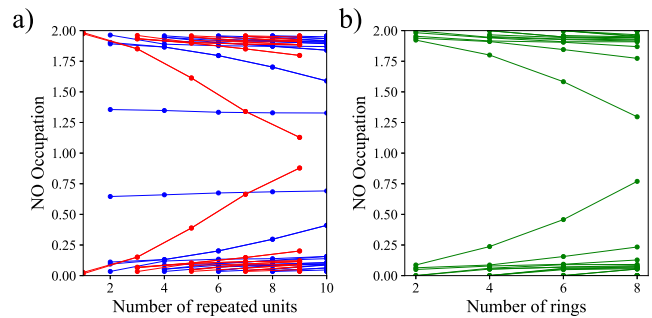}
\caption{
NO occupations for periodic (a) and molecular (b) acene chain calculations. In (a), the even-unit calculations are colored blue and the odd-unit calculations are colored red, to emphasize the differences in occupation trends between the two sets of calculations. Notably, the odd-unit calculations have a similar HONO and LUNO occupation trend as the molecular calculations, whereas the even-unit calculations have the same HONO and LUNO occupations across all calculations.}\label{Fig:Acene_occ}
\end{figure}

\begin{figure}[h]
\includegraphics[scale=0.09]{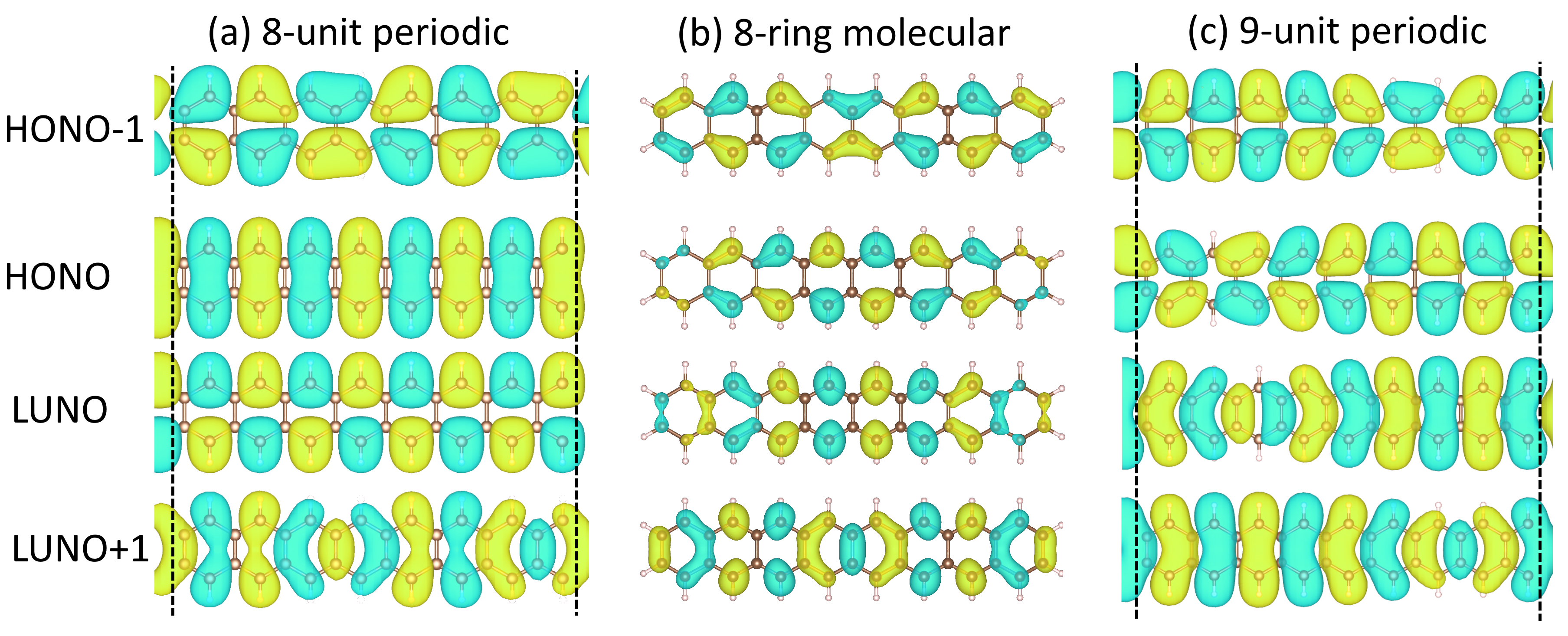}
\caption{Images of electron density for HONO-1, HONO, LUNO, and LUNO+1 for molecular (8-acene, (a)) and periodic (8-unit, (b), and 9-unit, (c)) calculations. The dashed lines in the structures for the periodic calculations represent the lattice boundary.}\label{Fig:Acene_NOs}
\end{figure}

NO occupations for periodic calculations are shown in Table~\ref{Tab:Acene_occ} and Figure~\ref{Fig:Acene_occ}, which show that the known polyradical nature~\cite{Bendikov2004, Jiang2008, Ni2018} of acene chains is only recovered by v2RDM for periodic boxes with an even number of unit cells included.  The  coupled cluster singles-doubles (CCSD) method with periodic boundary conditions only recovers static correlation in the 2 unit cell case.  Notably, the CCSD calculation does not exhibit significant static correlation in the 4 unit cell case. Multi-reference wavefunctions are needed to accurately capture static correlation, and CCSD, a single-reference method, fails to accurately capture static correlation in larger systems. This result is consistent with previous limitations with CCSD observed in the computation of finite acene chains~\cite{Gidofalvi2008}.  The highest-occupied natural orbitals (HONOs) and lowest-unoccupied natural orbitals (LUNOs) that display biradical character have a $180^\circ$ phase change between unit cells (\ref{Fig:Acene_NOs}), which explains the requirement for an even number of unit cells to achieve the biradical character. This also agrees with previous molecular (non-periodic) calculations suggesting this same parity~\cite{Hemmatiyan2019}. Although one unit cell contains only 4 carbon atoms and 2 hydrogen atoms, gamma-point calculations should be performed on a box containing two unit cells to accurately reflect molecular properties.

Electron densities for 8 ring molecular, 8-unit periodic, and 9-unit periodic calculations are included in Figure~\ref{Fig:Acene_NOs}. These images show clearly that the HONO-1 and LUNO+1 orbitals from the molecular calculation have even parity, while the HONO and LUNO have odd parity. Additionally, the electron density of the even parity orbitals remain unchanged in the 8-unit periodic calculation, and the odd parity orbitals remain unchanged in the 9-unit periodic calculation (after translation along the periodic axis). As a result, the occupations of the HONO and LUNO closely match those in the 9-unit periodic calculation, whereas the HONO-1 and LUNO+1 orbital occupations closely match those in the 8-unit periodic calculation (shown in Figure~\ref{Fig:Acene_occ}). Notably, each orbital in the 9-unit periodic calculation has a single "defect" which looks like the electron density along the edge of the molecular calculation. Because there are an odd number of repeating units in the unit cell, and gamma point calculations can only recover orbitals with an even number of antinodes, the defects effectively stretch an even number of antinodes across the unit cell. Finally, the symmetry of the 8-unit periodic calculation HONO and LUNO explain the early convergence of the occupation. Since these orbitals are just 4 copies of the 2-antinode pattern fully contained in 2 repeated units, these orbitals in particular stay the same for any even number of repeated units in the unit cell.

Earlier results using molecular DMRG~\cite{Hachmann2007} also indicate the polyradical behaviour of acene chains, but significant static correlation is only recovered for chain lengths longer than 6 rings, and the partial occupation levels only reach the values obtained here (1.35 HONO and 0.66 LUNO occupations for 6 unit cells) for chain lengths of 10 rings. Thus, we conclude that this periodic v2RDM method accurately captures strong correlation efficiently compared to DMRG methods, in that fewer atoms and orbitals are needed to observe strong correlation effects. Additionally, this method is generalizable to 2- or 3- dimensional systems (by summing over neighboring cells in 2 or 3 axes in Equation~\ref{Eq:periodic}), whereas DMRG methods are typically restricted to 1-dimensional systems.

\begin{figure}[h]
\includegraphics[scale=0.45]{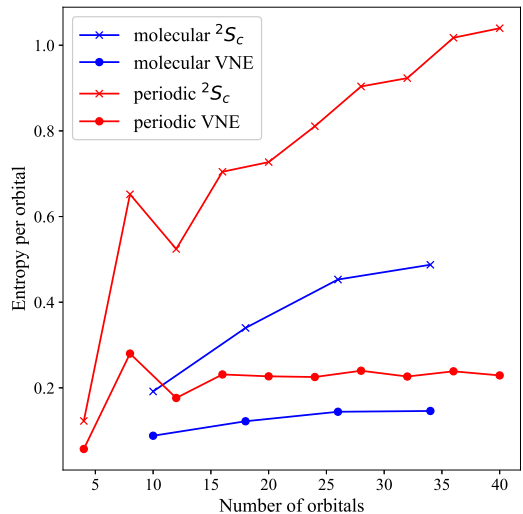}
\caption{1-electron Von Neumann entropy and connected entropy (${}^2S_{c}$) of molecular and periodic acene chains of various lengths. Periodic calculations have active spaces of [$4n, 4n$] where $n$ is the number of repeated units in the unit cell, and molecular calculations have active spaces of [$4n+2, 4n+2$] where $n$ is the number of rings in the calculation. The entropy of each calculation has been divided by the number of orbitals to get an intensive entropy measure that can be compared directly between calculations.}\label{Fig:Acene_entropy}
\end{figure}

For the both the periodic and molecular acene calculations, we include the 1-electron Von Neumann Entropy (1-VNE) and a 2-electron form of entropy of quantum entropy (we call the connected entropy) originally created by Prezhdo~\cite{Luzanov2007} in Figure~\ref{Fig:Acene_entropy}. The connected entropy is defined as:

\begin{equation}
\label{Eq:ConnectedEntropy}{}^2S_{c} = NS({}^1D) - S({}^2D)
\end{equation}
where $N$ is the number of electrons and $S(M)=-\textup{Tr}(M\ln{M})$ is the Von Neumann entropy. Both the 1-VNE and the connected entropy are extensive, and the 1-VNE quantifies the amount of bipartite correlation whereas the connected entropy quantifies the amount of tripartite correlation in the system. To compare these quantities between calculations, we divide each by the number of orbitals in the calculation to compare the intensive forms of the entropy. Because the all calculations performed use the complete $\pi$-space, these intensive quantities are directly comparable.

The intensive VNE is relatively constant for both molecular and periodic calculations, indicating that bipartite correlation is not increasing, despite the increase in correlation as longer chains are considered. In contrast, the connected entropy increases steadily as chain length increases, indicating an increase in tripartite correlation. Additionally, the periodic calculations have more of each type of entropy than the molecular calculations, for all chain lengths. As a result, the 2-unit periodic calculation recovers more correlation than any molecular calculation performed, which emphasizes the importance of including periodic boundary conditions to capture static correlation in extended systems.

\section{\label{Sec:Conc}Discussion and Conclusions}

We have generalized the variational 2-electron reduced density matrix (v2RDM) electronic structure theory, which incorporates periodic boundary conditions for extended structures. Using periodic boundary conditions can greatly simplify the computational complexity and accuracy of calculations by removing edge effects. Even gamma-point calculations, for which the electronic structure repeats with no phase change between unit cells, and therefore, does not account for long-range or low-frequency contributions to the wavefunction, can recover a large proportion of the correlation due to periodicity. The proposed periodic v2RDM method was shown to account for a large degree of static correlation due to the multireference nature of 2-RDMs and to correlate electrons between periodic cells. The ability to capture strong correlation in extended structures will prove beneficial in the study of many systems, including systems with extended $\pi$-systems, inorganic polymers, and materials, among others.

Additionally, we have confirmed previous work regarding the equilibrium bond length of the extended hydrogen chain. This is significant because we can efficiently recover an accurate estimate for the ground-state energy of the hydrogen chain system without performing expensive calculations with different chain lengths. We also showed that the periodic v2RDM method can capture the metal-to-insulator transition for the hydrogen chain, which Hartree-Fock, MP2, CISD, and CCSD calculations are unable to do.  The success of the v2RDM method is due to its accurate treatment of the strong correlation upon dissociation and its correct description of the periodic nature of the wavefunction.  With the application of this method to the acene chain, we have shown that parity effects require that two crystallographic unit cells are used for gamma-point electronic structure calculations, but we speculate that periodic calculations with more $k$-points can capture the parity with even a single unit cell.

Due to the ability to represent multireference systems, the 2-RDM methods can accurately capture strongly correlated materials without the exponential scaling of full configuration interaction, allowing the use of much larger active spaces than traditionally possible for such systems.  The combination of these attributes has the potential for more realistic descriptions of correlation-driven phenomena in extended $\pi$-systems, semiconductors, superconductors, organometallic polymers, as well as other materials.

\noindent {\bf Data Availability}: The data that support the findings of this study are available from the corresponding author upon reasonable request.

\begin{acknowledgments}

D.M. gratefully acknowledges the U.S. National Science Foundation Grant No. CHE-1565638 for support.

\end{acknowledgments}

\bibliography{PBC_paper}

\end{document}